\documentclass[prl,showpacs,twocolumn,unsortedaddress,floatfix,showkeys,aps,10pt]{revtex4-1}
\usepackage{physics}
\usepackage{subdepth}
\usepackage[version=4]{mhchem}
\usepackage{amsmath}
\usepackage{mathrsfs} 
\usepackage{amssymb}
\usepackage{graphicx}
\usepackage[normalem]{ulem}
\usepackage[unicode=true, pdfusetitle, bookmarks=true, bookmarksnumbered=false, bookmarksopen=false, breaklinks=true, pdfborder={0 0 0}, backref=false, colorlinks=true, linkcolor=blue, citecolor=blue, urlcolor=blue]{hyperref}
\allowdisplaybreaks
\newcommand{\smalldiv}{\raisebox{-0.2ex}{\resizebox{!}{1.6ex}{\kern0.1em/\kern0.1em }}}

\begin{document}

\title{(Un)physical consequences of ``Quantum Measurements of Time''}


\author{Will Cavendish}
\email{willcavendish@johnbellinstitute.org}
\affiliation{John Bell Institute for the Foundations of Physics, New York, NY 10003, United States}

\author{Siddhant Das}
\email{Siddhant.Das@physik.uni-muenchen.de}
\affiliation{Arnold Sommerfeld Center for Theoretical Physics (ASC), Fakult\"at f\"ur Physik, Ludwig-Maximilians-Universit\"at M\"unchen, Theresienstr.\ 37, D-80333 M\"unchen, Germany}

\author{Markus N\"oth}
\email{noeth@math.lmu.de}
\affiliation{Mathematisches Institut, Ludwig-Maximilians-Universit\"at M\"unchen, Theresienstr.\ 39, D-80333 M\"unchen, Germany}

\author{Ali Ayatollah Rafsanjani}
\email{aliayat@physics.sharif.edu}
\affiliation{Department of Physics, Sharif University of Technology, Tehran, Iran, School of Physics, Institute for Research in Fundamental Sciences (IPM), Tehran, Iran}

\date{August 8, 2024}

\maketitle

In \cite{MS}, Maccone and Sacha (hereafter, MS) claim ``to provide a general prescription for quantum measurements of the time at which an arbitrary event happens (the time of arrival being a specific instance).'' In this comment, we note that the empirical predictions of MS's ``Quantum Clock Proposal'' (QCP) are paradoxical when viewed as a solution to the quantum arrival-time problem (see \cite{Allcock1,Bogdan,MUGA1} for details on the standard formulation of the problem).

\begin{figure}[!ht]
    \centering
    \includegraphics[width=\columnwidth]{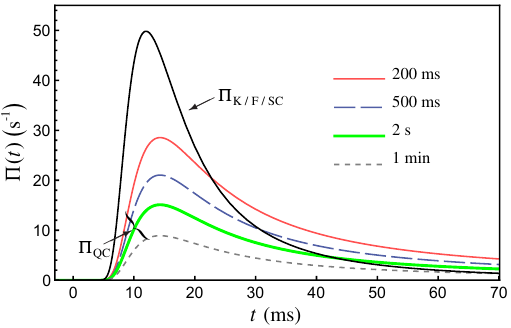}
    \caption{Time of arrival distributions for freely moving \ce{Ca^+} ions (mass \(\smash{m\approx 6.655\times 10^{-26}}\) kg) prepared in a Gaussian wave packet \cite[Eq.\ (17)]{Roncallo} of width \(\smash{\sigma_0=30\text{ nm}}\) and velocity \(\smash{p_0/m=5\text{ cm}/\text{s}}\approx 0.94\,\hbar/(m\sigma_0)\) for a flight distance \(\smash{|x_0|=1\text{ mm}}\). Kijowski's distribution (K) is, as is typical, nearly identical to the semi-classical (SC) and quantum flux (F) distributions. The quantum clock distribution (QC) is also drawn for select choices of the parameter \(T\) denominated in the legend.}
    \label{fig1}
\end{figure}

First, the empirical predictions of the QCP are dramatically different from those of other well-known proposals in the literature. Plotting the time-of-arrival (ToA) distributions for a free Gaussian wave packet with experimentally feasible parameters yields nearly indistinguishable distributions for Kijowski's distribution $\Pi_{\text{K}}$, the quantum flux distribution $\Pi_{\text{F}}$, and a semi-classical distribution $\Pi_{\text{SC}}$. The QC ToA distribution \(\Pi_{\text{QC}}\), on the other hand, is visibly different over a wide range of values of the ``regularization parameter'' \(T\) present in its definition, \cite[Eq.\ (6)]{MS} and \cite[Eq.\ (14)]{Roncallo}. Such differences are apparent even with parameters comparable to those examined by MS and Roncallo in \cite{Roncallo}. Indeed, redrawing Fig.\ 3 from \cite{Roncallo} with $\smash{p_0\approx 1\kern0.1em\hbar\smalldiv\sigma_0}$ rather than $\smash{p_0=7\kern0.1em\hbar\smalldiv\sigma_0}$ yields a graph similar to Fig.\ \ref{fig1}.

As Fig.\ \ref{fig1} shows, $\Pi_{\textrm{QC}}$ decreases pointwise with increasing $T$. In fact, letting $\smash{T\to \infty}$ as suggested in \cite{MS} typically causes $\Pi_{\textrm{QC}}$ to \emph{vanish}, not only for Gaussian wave packets but for \emph{all} wave functions \(\psi\) for which \(\smash{\braket{p}{\psi}\!|_{p=0}\ne 0}\). This is a dense set of wave functions that includes Gaussians. In particular, the denominator of $\Pi_{\text{QC}}$ diverges like \(\ln T \) \cite{dollard1969scattering}. This happens in \(\mathbb{R}^d\) for any \(d\) whenever the detector has codimension \(1\), e.g., for planar detection surfaces for \(\mathbb{R}^3\). Despite MS's insistence that this is a strength rather than a flaw of the QCP \cite{MS}, it means that the proposal is typically trivial and therefore empirically inadequate.

Fig.\ \ref{fig1} also shows that the QCP is empirically implausible even when $T$ is finite and is ``given by the total duration of the experiment'' \cite{Roncallo}. This is because the probability of arrival during some fixed interval $[t_1,t_2]$ depends on $T$ even for $\smash{T\gg t_2}$, i.e., it depends on the future time at which the experiment will conclude. While there may be an interpretation of the QCP that makes this seem less bizarre, it shows that \(\Pi_{\text{QC}}\) is not comparable to the ToA proposals surveyed in \cite{Roncallo}.

\begin{figure}
    \centering
    \includegraphics[width=\columnwidth]{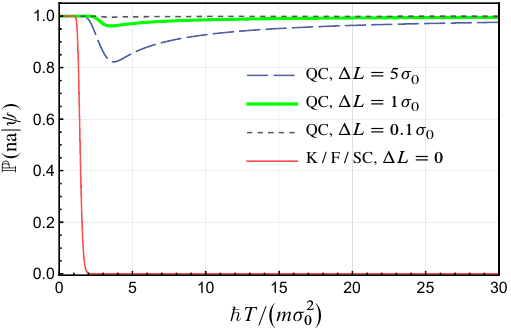}
    \caption{Non-detection probabilities for the Gaussian wave packet for \(\smash{|x_0|=10\kern0.1em\sigma_0}\), \(\smash{p_0=7\hbar/\sigma_0}\), and \(\smash{D=[-\Delta L/2,\Delta L/2]}\) vs.\ nondimensionalized cutoff time \(T\) for different proposals.}
    \label{NAComp}
\end{figure}

A final comment concerns MS's operator \(\Pi_{\text{na}}\) \cite[Eq.\ (3)]{MS}, which determines the \emph{non}-arrival probability
$$
\mathbb{P}_{\text{QC}}(\text{na}\kern0.1em|\kern0.1em\psi)= 1-\frac{1}{T}\int_{-T/2}^{T/2}dt\int_Ddx~|\psi(x,t)|^2.
$$
As Fig.\ \ref{NAComp} shows, this probability behaves very differently from \(\smash{\mathbb{P}_{\text{K / F / SC}}(\text{na}\,|\,\psi)= \int_T^{\infty}\!\!dt\,\Pi_{\text{\text{K / F / SC}}}(t)}\). Indeed, for any given $\psi$, the non-arrival probability for the F, K, and SC proposals decrease monotonically as a function of the cutoff $T$. This makes intuitive sense for laboratory arrival time experiments, because extending the duration of the experiment allows the particle to be detected over a longer period. In contrast, the QCP non-arrival probability approaches \(1\) for large \(T\). Applied to standard laboratory arrival time experiments, this would predict that a particle prepared in any state is almost certain \emph{not} to arrive if $T$ is sufficiently large. An elementary application of the long-time asymptotics \cite{dollard1969scattering} implies that this is the case for \emph{all} wave functions.

From the above considerations, it is evident that the QCP cannot be a solution to the arrival-time problem as is commonly understood.

\bibliography{thebibliography}

\begin{thebibliography}{6}%
\makeatletter
\providecommand \@ifxundefined [1]{%
 \@ifx{#1\undefined}
}%
\providecommand \@ifnum [1]{%
 \ifnum #1\expandafter \@firstoftwo
 \else \expandafter \@secondoftwo
 \fi
}%
\providecommand \@ifx [1]{%
 \ifx #1\expandafter \@firstoftwo
 \else \expandafter \@secondoftwo
 \fi
}%
\providecommand \natexlab [1]{#1}%
\providecommand \enquote  [1]{``#1''}%
\providecommand \bibnamefont  [1]{#1}%
\providecommand \bibfnamefont [1]{#1}%
\providecommand \citenamefont [1]{#1}%
\providecommand \href@noop [0]{\@secondoftwo}%
\providecommand \href [0]{\begingroup \@sanitize@url \@href}%
\providecommand \@href[1]{\@@startlink{#1}\@@href}%
\providecommand \@@href[1]{\endgroup#1\@@endlink}%
\providecommand \@sanitize@url [0]{\catcode `\\12\catcode `\$12\catcode
  `\&12\catcode `\#12\catcode `\^12\catcode `\_12\catcode `\%12\relax}%
\providecommand \@@startlink[1]{}%
\providecommand \@@endlink[0]{}%
\providecommand \url  [0]{\begingroup\@sanitize@url \@url }%
\providecommand \@url [1]{\endgroup\@href {#1}{\urlprefix }}%
\providecommand \urlprefix  [0]{URL }%
\providecommand \Eprint [0]{\href }%
\providecommand \doibase [0]{http://dx.doi.org/}%
\providecommand \selectlanguage [0]{\@gobble}%
\providecommand \bibinfo  [0]{\@secondoftwo}%
\providecommand \bibfield  [0]{\@secondoftwo}%
\providecommand \translation [1]{[#1]}%
\providecommand \BibitemOpen [0]{}%
\providecommand \bibitemStop [0]{}%
\providecommand \bibitemNoStop [0]{.\EOS\space}%
\providecommand \EOS [0]{\spacefactor3000\relax}%
\providecommand \BibitemShut  [1]{\csname bibitem#1\endcsname}%
\let\auto@bib@innerbib\@empty
\bibitem [{\citenamefont {Maccone}\ and\ \citenamefont {Sacha}(2020)}]{MS}%
  \BibitemOpen
  \bibfield  {author} {\bibinfo {author} {\bibfnamefont {L.}~\bibnamefont
  {Maccone}}\ and\ \bibinfo {author} {\bibfnamefont {K.}~\bibnamefont
  {Sacha}},\ }\href {\doibase 10.1103/PhysRevLett.124.110402} {\bibfield
  {journal} {\bibinfo  {journal} {Phys. Rev. Lett.}\ }\textbf {\bibinfo
  {volume} {124}},\ \bibinfo {pages} {110402} (\bibinfo {year}
  {2020})}\BibitemShut {NoStop}%
\bibitem [{\citenamefont {Allcock}(1969)}]{Allcock1}%
  \BibitemOpen
  \bibfield  {author} {\bibinfo {author} {\bibfnamefont {G.~R.}\ \bibnamefont
  {Allcock}},\ }\href {\doibase 10.1016/0003-4916(69)90251-6} {\bibfield
  {journal} {\bibinfo  {journal} {Ann. Phys.}\ }\textbf {\bibinfo {volume}
  {53}},\ \bibinfo {pages} {253} (\bibinfo {year} {1969})}\BibitemShut
  {NoStop}%
\bibitem [{\citenamefont {Mielnik}(1994)}]{Bogdan}%
  \BibitemOpen
  \bibfield  {author} {\bibinfo {author} {\bibfnamefont {B.}~\bibnamefont
  {Mielnik}},\ }\href {\doibase 10.1007/BF02057859} {\bibfield  {journal}
  {\bibinfo  {journal} {Found. Phys.}\ }\textbf {\bibinfo {volume} {24}},\
  \bibinfo {pages} {1113} (\bibinfo {year} {1994})}\BibitemShut {NoStop}%
\bibitem [{\citenamefont {Muga}\ and\ \citenamefont {Leavens}(2000)}]{MUGA1}%
  \BibitemOpen
  \bibfield  {author} {\bibinfo {author} {\bibfnamefont {J.~G.}\ \bibnamefont
  {Muga}}\ and\ \bibinfo {author} {\bibfnamefont {C.~R.}\ \bibnamefont
  {Leavens}},\ }\href {\doibase 10.1016/S0370-1573(00)00047-8} {\bibfield
  {journal} {\bibinfo  {journal} {Phys. Rep.}\ }\textbf {\bibinfo {volume}
  {338}},\ \bibinfo {pages} {353} (\bibinfo {year} {2000})}\BibitemShut
  {NoStop}%
\bibitem [{\citenamefont {Roncallo}\ \emph {et~al.}(2023)\citenamefont
  {Roncallo}, \citenamefont {Sacha},\ and\ \citenamefont {Maccone}}]{Roncallo}%
  \BibitemOpen
  \bibfield  {author} {\bibinfo {author} {\bibfnamefont {S.}~\bibnamefont
  {Roncallo}}, \bibinfo {author} {\bibfnamefont {K.}~\bibnamefont {Sacha}}, \
  and\ \bibinfo {author} {\bibfnamefont {L.}~\bibnamefont {Maccone}},\ }\href
  {\doibase 10.22331/q-2023-03-30-968} {\bibfield  {journal} {\bibinfo
  {journal} {{Quantum}}\ }\textbf {\bibinfo {volume} {7}},\ \bibinfo {pages}
  {968} (\bibinfo {year} {2023})}\BibitemShut {NoStop}%
\bibitem [{\citenamefont {Dollard}(1969)}]{dollard1969scattering}%
  \BibitemOpen
  \bibfield  {author} {\bibinfo {author} {\bibfnamefont {J.~D.}\ \bibnamefont
  {Dollard}},\ }\href {https://doi.org/10.1007/BF01661573} {\bibfield
  {journal} {\bibinfo  {journal} {Commun. Math. Phys.}\ }\textbf {\bibinfo
  {volume} {12}},\ \bibinfo {pages} {193} (\bibinfo {year} {1969})},\ \bibinfo
  {note} {{N.B.:} For \(\smash{\psi(\vb{x})\in L^2\big(\mathbb{R}^d\big)}\),
  Lemma 2 generalizes to
  $$\lim_{t\to\pm\infty}\left|\kern-0.15em\left|\,e^{-\,it\kern0.1em
  H_0/\hbar}\psi(\vb{x})-\exp(\frac{i \text{x}^2}{2
  \kern0.1em\tau})\frac{\tilde{\psi}\!\left(\vb{x}/\tau\right)}{(i\tau)^{d/2}}\,\right|\kern-0.15em\right|=0,$$
  where \(\smash{\tau=\hbar\,t/m}\), \(H_0\) is the free-particle Hamiltonian,
  and \(\hbar^{-\,d/2}\tilde{\psi}(\vb{p}/\hbar)\) is the momentum-space wave
  function \(\braket{\vb{p}}{\psi}\).}\BibitemShut {Stop}%
\end{thebibliography}%

\end{document}